\documentstyle[preprint,prl,aps]{revtex}

\begin{document}
\draft

\title{Transport Properties of a One-Dimensional Two-Component Quantum Liquid with Hyperbolic Interactions}

\author{Rudolf A.\ R\"{o}mer and Bill Sutherland}

\address{Physics Department, University of Utah, Salt Lake City, Utah 84112}

\date{February 28, 1994; printed \today}
%\date{\today}
\maketitle

\begin{abstract}
We present an investigation of the sinh-cosh (SC) interaction model with 
twisted boundary conditions.
We argue that, when unlike particles repel, the SC model may be usefully 
viewed as a Heisenberg-Ising fluid with moving Heisenberg-Ising spins.
We derive the Luttinger liquid relation for the stiffness and the 
susceptibility, both from conformal arguments, and directly from the 
integral equations. 
Finally, we investigate the opening and closing of the ground state gaps
for both SC and Heisenberg-Ising models, as the interaction strength
is varied.
\end{abstract}


\narrowtext
%\tighten

%%%%%%%%%%%%%%%%%%%%%%%%%%%%%%%%%%%%%%%%%%%%%%%%%%%%%%%%%%%%%%%%%%%%%%%%
%
% Introduction
%
%%%%%%%%%%%%%%%%%%%%%%%%%%%%%%%%%%%%%%%%%%%%%%%%%%%%%%%%%%%%%%%%%%%%%%%%

In Ref. \cite{sr93} and \cite{rs93}, to be called I and II in the following, 
we solved the integrable one-dimensional (1D) SC-model defined by the 
Hamiltonian
 \begin{equation}
H = - \frac{1}{2} \sum_{1\leq j \leq N} 
       \frac{\partial^{2}}{\partial x_{j}^{2}}
    + \sum_{1\leq j < k \leq N} v_{jk}(x_{j}-x_{k}).
\label{eqn-ham}
\end{equation}
The pair potential is given as
\begin{equation}
v_{jk}(x) = s(s+1)
 \left[
  \frac{1+\sigma_{j}\sigma_{k}}{2 {\rm sinh}^{2}(x)}
  -
  \frac{1-\sigma_{j}\sigma_{k}}{2 {\rm cosh}^{2}(x)}
 \right], \quad s>-1,
\label{eqn-pot}
\end{equation}
and the quantum number $\sigma=\pm 1$ distinguishes the two kinds of particles.
We may usefully think of it as either representing charge or spin.
For values of the interaction strength $s$ in the range $-1<s<0$, the system
exhibits two gapless excitation branches with different Fermi velocities as
does the repulsive 1D Hubbard model\cite{fk90}, and thus may be classified as 
a typical two-component 1D Luttinger liquid\cite{h8x}.
The asymptotic behavior of the correlation functions is given 
by finite-size arguments of conformal field theory.
A Wiener-Hopf type calculation\cite{rs93} shows that the spin-spin 
part of the dressed charge matrix is essentially identical to the dressed 
charge scalar in the Heisenberg-Ising (H-I) model\cite{yy}.  

In this Letter, we will further explore the connection of the SC model
with the H-I model by examining the response of the system to a flux 
$\Phi$.
The addition of a flux is compatible with integrability and allows the
study of the transport properties by an adiabatic variation of $\Phi$. 
For the H-I model, this has already been done\cite{ss90,yf90} for the
interaction strength range $-1\leq\Delta\leq1$. We will show that the spin
degrees of freedom of the SC model for $0>s>-1$ may be usefully 
viewed as a H-I model with moving H-I spins. The presence of the translational
degrees of freedom will simply renormalize the spin-spin coupling.

We thus restrict ourselves in what follows to the unbound case $-1<s<0$, such 
that there are two gapless excitations corresponding to a particle-hole 
and a two spin-wave continuum with excitation velocities $v$ and $v_s$,
respectively. 
Let us then modify the Bethe ansatz equations of Eq.(II.7) by threading
them with a flux $\Phi$.
We have two coupled equations for $N$ particles with pseudo-momenta
${\bf k}=(k_{1}, \ldots, k_{N})$ and $M$ spin waves with
rapidities $\mbox{\boldmath $\lambda$}=(\lambda_{1}, \ldots, \lambda_{M})$
on a ring of length $L$.
The energy of a given state is $E({\bf k})=\frac{1}{2}\sum_{j=1}^{N} k_{j}^2$ 
and the total momentum is $P({\bf k})=\sum_{j=1}^{N} k_{j}$.
Boosting the system by $\Phi$ will accelerate the two kinds of particles in 
opposite directions due to the two components being of equal but 
{\em opposite} charge. 
Therefore, we have no center-of-mass motion and $P=0$.
The energy of a given state will change as a function of 
$\Phi$, and the energy shift of the ground state may be written as
$\Delta E_0(\Phi)\equiv E_0(\Phi)-E_0(0)\equiv D\Phi^2/2L + O(\Phi^4)$, 
where $D$ is called the stiffness constant and can be specified by 
perturbation arguments for $\Phi$ up to $\pi$\cite{ss90}.
Note that since we do not have any center-of-mass motion, we can call
$D$ either spin or charge stiffness depending on what interpretation of
$\sigma$ we adopt. We choose the spin language for comparison with the H-I 
model. 
However, the charge interpretation is probably more natural to describe 
transport properties.
We furthermore caution the reader that the term charge stiffness has been
previously used in lattice models to describe center-of-mass motion. 

The twisted Bethe ansatz equations are given by
\begin{mathletters}
\label{eqn-bakl}
\begin{equation}
-L k_{j} 
 = 2\pi I_{j}(k_{j}) 
 - \frac{M}{N} \Phi 
 + \sum_{a=1}^{M}\theta_{0,-1}(k_{j}-\lambda_{a})
 + \sum_{l=1}^{N}\theta_{0,0}(k_{j} - k_{l}), 
\label{eqn-bak}
\end{equation}
\begin{equation}
0 
 = 2\pi J_{a}(\lambda_{a}) 
 + \Phi 
 + \sum_{b=1}^{M}\theta_{-1,-1}(\lambda_{a}-\lambda_{b})
 + \sum_{j=1}^{N}\theta_{0,-1}(\lambda_{a}- k_{j}).
\label{eqn-bal}
\end{equation}
\end{mathletters}
The two-body phase shifts for particle-particle, particle-spin wave and 
spin wave-spin
wave scattering, $\theta_{0,0}(k)$, $\theta_{0,-1}(k)$ and 
$\theta_{-1,-1}(k)$ respectively, have been given in I.
The particle quantum numbers $I_{j}$ and the spin-wave quantum numbers
$J_{a}$ are integers or half-odd integers depending on the parities 
of $N$, $M$ as well as on the particle statistics\cite{rs93}. 
For simplicity, we use bosonic selection rules, although a purely
fermionic or a mixed bose-fermi system may be studied along
similar lines. In the ground state of the bosonic system, we have
\begin{eqnarray}
I_1, I_2, \ldots, I_N &=
 & -\frac{(N-1)}{2}, -\frac{(N-3)}{2}, \ldots, \frac{(N-1)}{2}, \nonumber \\
J_1, J_2, \ldots, J_M &=
 &-\frac{(M-1)}{2}, -\frac{(M-3)}{2}, \ldots, \frac{(M-1)}{2},
\end{eqnarray}
for both N and M even. 

We start with some general considerations.
Let us denote by $E_{\{{\bf I},{\bf J}\}}(\Phi)$ the energy of a state 
specified by the $\Phi=0$ set of quantum numbers $\{{\bf I},{\bf J}\}$.
We then adiabatically turn on the flux until we return to our initial
state. The energy will also return to its initial value, although, it
may return sooner; so, the period of the wave function
will be an integer multiple of the period of the energy.
We can define a topological winding number $n$ to be the number of times
the flux $\Phi$ increases by $2\pi$ before the state returnes to its
initial value.
As Sutherland and Shastry have shown, the ground state winding number 
of the H-I model with $S_z=0$ in the parameter range 
$-1<-\cos(\mu)\equiv\Delta<1$ is $2$, implying charge carriers with half
the quantum of charge, {\em except}
at isolated points $\Delta=\cos(\pi/Q)$, where $M\geq Q\geq 2$ is an integer.
In particular, at $\Delta=0$, the free particle wave function has
periodicity $2\pi N_{HI}$, where $N_{HI}$ is the number of H-I sites,
implying free acceleration in the thermodynamic limit.

We now note the following important fact:
Choosing $\mu\equiv -\pi s$, 
the spin wave-spin wave phase shift $\theta_{-1,-1}$ is 
identical to the spin-spin phase shift in the H-I model, 
and we may rewrite the equation for the rapidities as
\begin{equation}
 N \bar{\theta}_{0,-1}(\lambda_{a},\mu) \equiv
 N \sum_{j=1}^{N}\theta_{0,-1}(\lambda_{a}- k_{j},\mu)/N
 = 2\pi J_{a}(\lambda_{a}) 
  + \Phi 
  + \sum_{b=1}^{M}\theta_{-1,-1}(\lambda_{a}-\lambda_{b},\mu).
\label{eqn-balHIM}
\end{equation}
which nearly is identical to the Bethe Ansatz equation of the H-I model, as can
be readily seen when we use the standard transformation for the H-I momenta
$p= f(\alpha,\mu)$. We then merely have to identify $\alpha\equiv\pi\lambda$. 
The sole effect of the pseudo-momenta ${\bf k}$ is an averaging on the left 
hand side. 
Let us now restrict ourselves in what follows to the neutral (spin zero)
sector such that we have $M$ particles with $\sigma=-1$ and $M$ particles 
with $\sigma=+1$ for a total of $N=2 M$.
Then, a discussion of the behavior of the rapidities 
$\mbox{\boldmath $\lambda$}$ for
varying $\Phi$ exactly mimics the discussion of the H-I momenta $p$ in
Ref.\cite{ss90} at $S_z=0$: As long as $|\Phi|\leq 2\pi (s+1)$, all $\lambda$'s
stay on the real axis. At $\Phi=2\pi (s+1)$, the largest root
$\lambda_M$ goes to infinity.
For $\Phi$ increasing beyond this point, $\lambda_M$
will reappear from infinity as $i\pi+\gamma_1$ until exactly at
$\Phi=2\pi$, $\lambda_M=i\pi$ ($\gamma_1=0$) and the remaining $M-1$ rapidities
have redistributed themselves symmetrically around $0$ on the real axis.
However, as mentioned above, this behavior is different at the threshold values 
$s=(1-Q)/Q$.
The momenta ${\bf k}$ are always real and distributed about the origin.
Eq.(\ref{eqn-bal}) simplifies at $\Phi=2\pi (s+1)$ (and thus $\lambda_M=\infty$)  
and is in fact just the equation for $M-1$ rapidities in the ground state.
So as in Ref.\cite{ss90} using simple thermodynamical arguments, we may write
\begin{equation}
\Delta E_0(2\pi (s+1)) = E_0(N,M-1)-E_0(N,M) = 1/2L \cdot \chi^{-1},
\label{eqn-de}
\end{equation}
where $\chi$ is the susceptibility.
Comparing with the definition of the stiffness constant $D$, we find
$D= \chi^{-1}/4 \pi^2 (s+1)^2$.

On the other hand, we can read off the finite-size energy corrections for
the SC model, and then finite-size arguments of conformal field theory give
an expression for $\Delta E_0(2\pi(s+1))$ in
terms of the conformal weights, the dressed charge matrix $\Xi$ and the 
spin wave velocity $v_s$. The neutral sector dressed 
charge matrix is given in Eq.(II.35) and thus we have
$\chi^{-1}= 2\pi v_s (s+1)$.
We may therefore express the stiffness $D$ in terms of the spin 
wave velocity as
\begin{equation}
D=v_s/2\pi (s+1).
\label{eqn-d}
\end{equation}
We emphasize that this formula for $D$ is true also for a system of purely 
fermionic particles.
Shastry and Sutherland \cite{ss90} have given an exact formula for the stiffness
constant in the H-I model, by using the known expression for the H-I model spin wave
velocity $v_s= \pi \sin(\mu) / \mu$ \cite{ycg}. No such expression is
known for the SC model and we can only give $v_s$ as
\begin{equation}
v_s = \frac{1}{2\pi}
\frac{ \int_{-B}^{B} e^{-\pi k/2s} \epsilon'(k) dk }{\int_{-B}^{B} e^{-\pi k/2s} \rho(k) dk }.
\label{eqn-vs}
\end{equation}
Here, we use the definitions of II, Section II.
However, written in terms of spin velocities the stiffness formulas are 
identical and only the values of the respective spin wave velocities are 
different. Thus the presence of the translational degrees of freedom in the
SC model simply renormalizes the spin-wave velocity.

We have iterated the Bethe Ansatz equations (\ref{eqn-bakl}) in the neutral
sector for reasonably 
large systems and density $d\equiv N/L =1/2$ as a function of $\Phi$. 
By our correspondence between the H-I model, and the spin wave
part of the SC model, we expect {\em free spin waves} at $s=-1/2$.
In the thermodynamic limit, we would thus expect the periodicity of the ground 
state energy to be infinite. For a finite system, this will be reduced to
a periodicity that scales with the system size.
For the SC model we have indeed found that at $s=-1/2$ the periodicity of the
ground state energy is given as $2\pi N$. 
We may then speak of $s\rightarrow -1^+$ as the ferromagnetic critical point 
and $s\rightarrow 0^-$ as the antiferromagnetic critical point of the SC model.
In Fig.(\ref{fig-1}) we show the full spectrum of low-lying states with
zero momentum at $s=-1/2$ for $L=12$, $N=6$ and $M=3$. 
The ground state curve is emphasized and its periodicity is $6\cdot2\pi$. 

Note that at $\Phi=2\pi$ there is a level 
crossing between ground state and first exited state in Fig.(\ref{fig-1}).
When the interaction strength changes from $s=-1/2$, immediately 
a gap opens between the ground state and first exited state. 
Just as in the H-I model the periodicity is reduced to $4\pi$.
Note that a perturbation theory argument can not describe this behavior.
Fig.(\ref{fig-2}) shows the behavior of the ground state energy variation 
$L [1-E(\Phi)/E(2\pi)]$ for $s=-1/3$ near $\Phi=2\pi$ for different lattice 
sizes.
The rounding is well pronounced and does not vanish as we increase the size.

Thus the behavior of the low-lying states in the SC and H-I models is
qualitatively the same, up to the renormalization of quantities such as the 
spin wave velocity $v_s$.
Let us briefly describe the behavior of the gaps in the H-I model,
keeping in mind the correspondence $\mu=-\pi s$.
Increasing $\mu$ beyond $\pi /2$ ($\Delta= 0$), we see that the gap continues 
to widen up to a maximum value at $\mu\sim 7\pi /12$ ($\Delta\sim 0.26$). 
It then closes up again exactly at $\mu=2\pi /3$ ($\Delta= 1/2$). 
As has been noted before, this value of $\mu$ coincides with the appearance 
of a $Q=3$ string\cite{yy}. 
Further increase of $\mu$ again opens, and then closes the gap at the threshold 
for the next-longer $Q=4$ string. This behavior continues, and the threshold
values accumulate as $\mu\rightarrow\pi$ ($\Delta\rightarrow 1$). 
In Fig.(\ref{fig-3}), we show the ground state and the first exited state
of the H-I model on a ring of $N_{HI}=12$. Note that due to the finite
size of the ring, we can only observe strings up to length $Q=6$.
We will present a more detailed finite-size study of the behavior of the
gaps in H-I and SC model in another publication.
We only mention that for fixed $\mu$ the gap scales with the system size
as a negative power of $N_{HI}$, with variable exponent depending on the 
coupling constant $\mu$.

The stiffness constant $D$ is the curvature of the ground 
state energy $E_0(\Phi)$ as a function of $\Phi$. In
Fig.(\ref{fig-4}), we show $D$ for systems of $12$, $24$ and $32$ lattice
sites. We also show the behavior of $D$ as given by Eq.(\ref{eqn-d}).
As $s\rightarrow 0^{-}$, the spin wave velocity approaches the velocity
of a non-interacting single-component model, i.e. $v_s\rightarrow \pi d/2$ 
\cite{rs93}. 
Thus $D$ approaches the non-zero value $1/8$ which is compatible with the
result of Ref.\cite{ss90}.
Furthermore,  the SC model exhibits a gap for $s>0$ and so $D$ is zero. Thus
$D$ exhibits a jump discontinuity at $s=0$ just as in the H-I model for
$\Delta=-1$.

Note that Eq.(\ref{eqn-d}) may also be written as $D\chi^{-1}=v_s^2$. This is
nothing but the Luttinger relation for the spin wave excitations \cite{h8x}. 
Let us briefly explain how to derive this formula without using
arguments of conformal field theory.
In the thermodynamic limit, we convert Eq.(\ref{eqn-bakl}) into a set of
coupled integral equations as in Eq.(II.10). Here $\rho(k)$ and 
$\sigma(\lambda)$ are the distribution functions of particles and down-spins,
respectively. The density $d$ and the magnetization ${\cal M}$ are then
given parametrically in terms of the integral limits $B$ and $C$.
We now use an iteration scheme, i.e., 
first, for $B=\infty$ and ${\cal M}=0$, i.e. at half-filling and zero 
magnetization, we calculate $\rho(k)\equiv\rho_0(k)$.
We then use this $\rho_0(k)$ in the equation for $\sigma(\lambda)$ with
$C$ finite and $B$ nearly $\infty$.
Finally, we use this $\sigma(\lambda)$ to calculate $\rho(k)$ and thus the 
effect on the momenta and the energy. Since we are only interested in the 
leading order correction terms, we may stop. 
Then the corrections to the energy are
\begin{eqnarray}
\frac{\Delta E}{L} 
&\equiv &\frac{1}{2} \left[ D (\frac{\Phi}{L})^2 + \chi^{-1} {\cal M}^2 \right]
 \nonumber \\
&=      &\frac{v_s}{4\pi (1+s)} \left[ (\frac{\Phi}{L})^2 + 
                   [2\pi (1+s)]^2 {\cal M}^2 \right],
\end{eqnarray} 
where ${\cal M}= \case{1}{2} d(1-2 M/N)$.
A complete account of this calculation can be found in Ref.\cite{rs94pla}.

The derivation of the Luttinger relation uses integral equations and as
such is valid in the thermodynamical limit. 
Most of the other results given above have been derived using 
Eq.(\ref{eqn-bakl}). 
These equations, however, have been derived by the
asymptotic Bethe Ansatz (AsymBA). This method is only correct 
in the thermodynamic limit\cite{su94}. 
Thus all our finite-size results should exhibit correction terms. 
From the hyperbolic form of the pair potential (\ref{eqn-pot}), we may expect 
these corrections to be exponentially small in $L$.
Indeed, a log-log plot of the ground state energy versus $L$ at fixed 
interaction strength shows a simple power law behavior already for $L\geq 6$.
Thus the $L\rightarrow\infty$ behavior of the finite-size Bethe-Ansatz 
equations for the SC model does not seem to differ in any significant respect 
from usual finite size behavior for short ranged models. 
This further supports our use of the AsymBA in the present study. 
     
In conclusion, we have shown that the SC model exhibits all the rich structure
of the H-I model for $-1<s<0$. In particular, there is a Luttinger relation
for the spin waves just as in the H-I model, that can be derived from
(i) conformal arguments, (ii) an exact calculation in the thermodynamic limit 
and (iii) is furthermore supported by numerical results for finite systems.
Thus this yields credibility to both the conformal and the Luttinger
approach in models solved by the AsymBA.
Finally, we have reported an interesting behavior of the gaps in H-I and 
SC models.

%\acknowledgements

R.A.R.\ gratefully acknowledges partial support by the Germanistic Society
of America.

% figures

\begin{figure}
  \caption{The low-lying states for the SC model at $L=12$, $N=6$ and $M=3$.
   The bold curve corresponds to the ground state and the winding number 
   is $n=6=N$.
   Note the various level crossing in this free spin wave case, especially
   the crossing of ground state and first exited state at 
   $\Phi=2\protect\pi$.
  \label{fig-1}}
\end{figure}

\begin{figure}
  \caption{Plot of the ground state energy variation $L [1-E(\Phi)/E(2\pi)]$ 
   for the SC-model at $s=-1/3$ for $L=12,20$ and $28$. 
  \label{fig-2}}
\end{figure}

\begin{figure}
  \caption{The charge stiffness $D(s)$ for the SC model. The dashed curves 
   correspond to $L=12$, $24$ and $32$ and converge to $D(0)=1/8$ at 
   $s\rightarrow 0^{-}$.
   The solid curve comes from Eq.(\protect\ref{eqn-d}), which can be derived by
   conformal methods or from the Luttinger relation.
   (Note that as $s\rightarrow 0^{-}$, the solid curve does not converge
   to $1/8$. This is due to a buildup of numerical errors in the
   integration routine.)
  \label{fig-3}}
\end{figure}

\begin{figure}
  \caption{Energy of the ground state and first exited state and their 
   difference in the H-I model for $N_{HI}=12$.
   Note the closing of the gap at 
   $\protect\Delta=\protect\cos(\protect\pi/Q)$ for 
   $Q=2,3,4,5$.   
  \label{fig-4}}
\end{figure}
\end{document}